\documentclass[letterpaper]{ptephy_v1}

\usepackage{amsmath}
\usepackage{graphics}
\usepackage{url}

\usepackage{color}

\def\edth{{\rlap{$\partial$}\raise0.3em\hbox{$-$}}}
\newcommand{\bea}{\begin{eqnarray}}
\newcommand{\eea}{\end{eqnarray}}

\begin{document}

\title{The detection of quasinormal mode with $a/M \sim 0.95$ would
prove a sphere $99\%$ soaking in the ergoregion of the Kerr space-time}


\author{Hiroyuki Nakano, Takashi Nakamura and Takahiro Tanaka}

\address{Department of Physics, Kyoto University, Kyoto 606-8502, Japan}

\begin{abstract}
Recent numerical relativity simulations of mergers of binary black holes suggest
that the maximum final value of $a/M$ is $\sim 0.95$ for
the coalescence of two equal mass black holes
with aligned spins of the same magnitude $a/M=0.994$
which is close to the upper limit $a/M=0.998$ of accretion spin-up shown by Thorne.
Using the WKB method, we suggest that if quasinormal modes
with $a/M \sim 0.95$ are detected
by the second generation gravitational wave detectors,
we could confirm the strong gravity space-time based on Einstein's general relativity
up to $1.33M$ which is only $\sim 1.014$ times
the event horizon radius and within the ergoregion. 
One more message about black hole geometry is expected here.
If the quasinormal mode is different from that of general relativity,
we need to find the true theory of gravity which deviates 
from general relativity only near the black hole horizon.
\end{abstract}

\subjectindex{E31, E02, E01, E38}

\maketitle

\section{Introduction}

The Kerr space-time~\cite{Kerr:1963ud}
which describes a rotating black hole (BH)
is very interesting not only in astrophysics
but also in mathematics and physics.
In particular, the existence of the ergoregion induces
various phenomena specific to strongly curved space-time,
such as the Penrose process~\cite{Penrose:1969pc}, 
by which we can extract the rotational energy of BH.

To confirm the space-time described by the Kerr BH using
the second generation gravitational wave detectors such as Advanced LIGO
(aLIGO)~\cite{TheLIGOScientific:2014jea}, 
Advanced Virgo (AdV)~\cite{TheVirgo:2014hva}
and KAGRA~\cite{Somiya:2011np,Aso:2013eba},
we discuss gravitational waves of the quasinormal modes (QNMs)
which are the unique signature of the gravitational wave emission from the BH.
We expect that the gravitational waves are emitted
when a BH is formed after the merger of compact objects,
and for example, the possible detection rate of BH-BH mergers
has been discussed in our previous paper~\cite{Nakamura:2016gri}.

It would be interesting here to observe that
there may be a restriction on the spin of BHs astrophysically.
From the mass formula of BH~\cite{MTW:1973},
we have the gravitational mass of $M$ as
\begin{equation}
 M^2 = \frac{2M_{\rm irr}^2}{1+\sqrt{1-q^2}} \,,
\end{equation}
where $q=a/M$ with the BH's specific angular momentum $a$, 
and $M_{\rm irr}$ is the irreducible mass of the Kerr BH which is related
to the area of the event horizon $A$ as $A=16\pi M_{\rm irr}^2$.
For a single BH, Thorne~\cite{Thorne:1974ve} showed that the maximum value
of $q_{\rm max}$ is $\sim 0.998$ since the radiation emitted
by the accretion disk carries the angular momentum to prevent
the BH from reaching the extremal limit, $q=1$.
Now, let us consider the merger of two equal mass Kerr BHs of mass
$M$ with the maximum value of aligned spins $q=q_{\rm max}$, 
which results in a single BH with the final mass $M_f$ and $q=q_f$.
Since the area of the horizon should increase~\cite{Hawking:1971vc}
(see also Ref.~\cite{Bardeen:1973gs}), we have
\begin{equation}
 {1+\sqrt{1-q_f^2}}\geq \frac{2M^2}{M_f^2}(1+\sqrt{1-q_{\rm max}^2})=\frac{2.126M^2}{M_f^2} \,.
\end{equation}
This means that if $M_f\geq 1.46M$, $q_f=1$ is possible in
contrast to the accretion spin-up model by Thorne~\cite{Thorne:1974ve}.
However, even in a recent numerical relativity simulation of a binary BH merger
with equal mass $M_1=M_2=M$ and aligned equal spins with $q_1=q_2=0.994$
by Scheel et al.~\cite{Scheel:2014ina}, $q_f$ is $\sim 0.95$. 
Therefore, we restrict our study up to $q_f=0.97$ here.

In our previous paper~\cite{Nakamura:2016gri},
we have presented a method to claim
how close to the event horizon of a BH we actually {\it see}
by gravitational wave detection of the QNMs.
In that paper, our focus was not in determining the QNM
frequencies accurately at all. 
Instead, we used the known accurate numerical results of the complex QNM frequencies
and the separation constant $\lambda$ in the Teukolsky
equation~\cite{Teukolsky:1973ha}, 
to suggest which part of strong gravity space-time 
is confirmed by the detection of QNMs.
In the present paper, using the same approach, we discuss further 
whether we can reach the confirmation of the space-time region 
within the ergoregion by the QNM gravitational wave detection.

This paper is organized as follows. In \S~2, we will argue our method briefly.
The results are given is \S~3 and \S~4 is devoted to discussions. 
We use the geometric unit system, where $G=c=1$ in this paper.

\section{Approach}

In the Boyer-Lindquist coordinates, the Kerr metric is given by
\bea
 ds^2 &=& 
 - \left( 1 - \frac{2 M r}{\Sigma} \right)  dt^2 
 - \frac{4 M a r ~{\rm{sin}^2 \theta} }{\Sigma} dt d\phi 
 + \frac{\Sigma}{\Delta} dr^2 + \Sigma d\theta^2
\cr &&
 + \left( r^2 + a^2 + \frac{2 M a^2 r}{\Sigma} \sin^2 \theta \right)
 \sin^2 \theta d\phi^2 \,,
\eea
where $M$ and $a$ are the mass and the spin parameter, respectively, 
$\Sigma = r^2 + a^2 \cos^2 \theta$ and $\Delta = r^2 - 2 M r + a^2$. 
We summarize here three characteristic radii
in this Kerr space-time.
The event horizon is located at
\bea
 r_{+}=M \left(1 + \sqrt{1-q^2} \right) \,,
\eea
and the inner light ring radius~\cite{Bardeen:1972fi} is at
\bea
 r_{\rm lr}= 2M \left\{1+\cos\left[\frac{2}{3}
 \cos^{-1}\left(-q\right)\right]\right\} \,,
\eea
which is evaluated in the equatorial ($\theta=\pi/2$) plane.
In Ref.~\cite{Yang:2012he} (and references therein),
the relation between this light ring orbit
and the QNM frequencies in the high frequency regime has been discussed.
Also, the ergosurface is given by
\bea
r_{\rm ergo}(\theta) = M \left(1 + \sqrt{1-q^2\cos^2 \theta} \right) \,,
\eea
and we denote the equatorial radius of the ergoregion as
\bea
 r_{\rm ergo} = 2M \,.
\eea

In the previous paper~\cite{Nakamura:2016gri},
we have extended the physical picture of QNMs 
by Schutz and Will~\cite{Schutz:1985zz}
for the Schwarzschild space-time
via the WKB method to the Kerr space-time.
Given the radial wave equation,
\bea
 \frac{d^2X}{dr^{*2}} + \left( \omega^2-V_{\rm D} \right) X = 0 \,,
 \label{eq:radial}
\eea
where $dr^*/dr = (r^2+a^2)/\Delta$
and $V_{\rm D}$ is a potential which is obtained from the potential 
of the Teukolsky radial equation~\cite{Teukolsky:1973ha}.
Here, we approximate the potential
by the expansion near the radius at its peak location $r^*_0$ as
\begin{eqnarray}
 V_{\rm D}(r^*) = V_{\rm D}(r^*_0)
 +\frac{1}{2} \left. \frac{d^2V_{\rm D}}{dr^{*2}}
 \right|_{r^*=r^*_0}(r^*-r^*_0)^2 \,.
\end{eqnarray}
Here, $r^*_0$ is related to the peak location
in the Boyer-Lindquist radial coordinate as
\bea
 r^*_0 = r_0 + \frac{2}{r_+ - r_-}
 \left[ r_+ \ln\left(\frac{r_0 - r_+}{2}\right)
 - r_- \ln\left(\frac{r_0 - r_-}{2}\right)\right]
\,,
\eea
where $r_- = M(1-\sqrt{1-q^2})$.
In practice, we determine $r^*_0$
by evaluating the peak location of the absolute value
of the potential in Eq.~\eqref{eq:radial}.
In the appendix of Ref.~\cite{Nakamura:2016gri}, we have also discussed
the location of $r^*$ which satisfies $dV/dr^*=0$ in the complex radius plane,
and then read off the effective peak radius from the real part
of the complex radius, to find no significant difference between two radii. 

Then, the QNM frequencies are derived as
\bea
 \omega^2 = (\omega_r+i\omega_i)^2
 = V_{\rm D}(r^*_0)-i\sqrt{-\frac{1}{2}
 \left. \frac{d^2V_{\rm D}}{dr^{*2}} \right|_{r^*=r^*_0}} \,,
\label{eq:WKBomega}
\eea
in the leading order WKB analysis.

Here, in Ref.~\cite{Nakamura:2016gri},
we used the WKB approximation to determine QNMs
by using the spatial positions of the maximum absolute values
of the Sasaki-Nakamura potential
$V_{\rm SN}$~\cite{Sasaki:1981kj,Sasaki:1981sx,Nakamura:1981kk}
and the Detweiler potential $V_{\rm D}$~\cite{Detweiler:1977gy}
up to $q=0.8$ since the remnant spin $q_f\sim 0.7$ is expected from
the results of numerical relativity for the merger of equal mass BHs
with $q_1=q_2=0$~\cite{Pretorius:2005gq, Campanelli:2005dd, Baker:2005vv}.
In this paper, we consider up to $q_f=0.97$ since a recent result
of numerical relativity for $q_1=q_2 \sim q_{\rm max}$~\cite{Scheel:2014ina}
yields $q_f\sim 0.95$.

Also, in this paper, we treat the Detweiler potential~\cite{Detweiler:1977gy}
(the $(-+)$ case in Ref.~\cite{Nakamura:2016gri})
as the potential $V_{\rm D}$. This is because the $(-+)$ potential has less wavy shape
and looks most suitable compared with the other cases for the WKB analysis
(see Figs.~1 and 5 of Ref.~\cite{Nakamura:2016gri}).
We do not repeat how to derive $V_{\rm D}$ since the details are
written in Sec.~4 of Ref.~\cite{Nakamura:2016gri}.  

In the following, we focus only on the ($\ell=2,\,m=2$) mode.
This is because the ($\ell=2,\,m=2$) QNM is dominant
in numerical relativity simulations of binary BH mergers
(see e.g., Ref.~\cite{London:2014cma})
even in the case of the remnant spin $q_f \sim 0.95$~\cite{Scheel:2014ina}
(we can check the behavior by using the waveforms
in ``SXS Gravitational Waveform Database''~\cite{SXSwaveforms}).

\section{Results}

\begin{figure}[!ht]
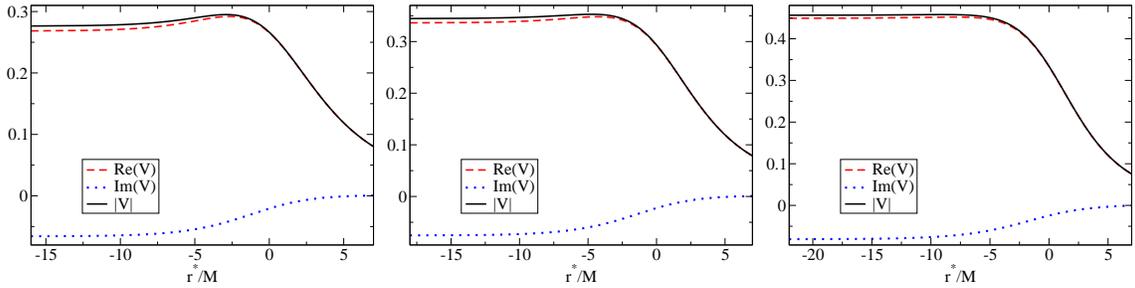

\begin{center}
 \includegraphics[width=0.32\textwidth,clip=true]{./Vmp07}
 \includegraphics[width=0.32\textwidth,clip=true]{./Vmp08}
 \includegraphics[width=0.32\textwidth,clip=true]{./Vmp09}
\end{center}
 \caption{${\rm Re}(V_{\rm D})$, ${\rm Im}(V_{\rm D})$ and $|V_{\rm D}|$ 
 with ($\ell=2,\,m=2$)
 for $q=0.7$ (left), $0.8$ (center) and $0.9$ (right)
 with each QNM frequency
 as a function of $r^*/M$ where we set $M=1$.
 The contribution of the imaginary part is small.}
 \label{fig:Vmp070809}
\end{figure}

\begin{figure}[!ht]
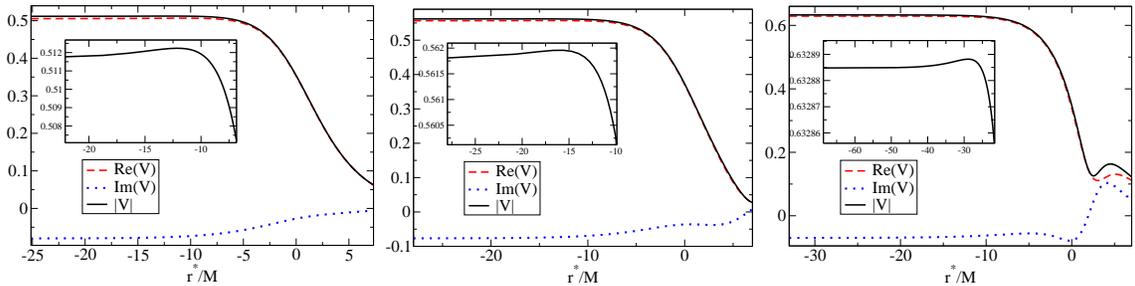

\begin{center}
 \includegraphics[width=0.32\textwidth,clip=true]{./Vmp093}
 \includegraphics[width=0.32\textwidth,clip=true]{./Vmp095}
 \includegraphics[width=0.32\textwidth,clip=true]{./Vmp097}
\end{center}
 \caption{${\rm Re}(V_{\rm D})$, ${\rm Im}(V_{\rm D})$ and $|V_{\rm D}|$ 
 with ($\ell=2,\,m=2$)
 for $q=0.93$ (left), $0.95$ (center) and $0.97$ (right)
 with each QNM frequency
 as a function of $r^*/M$ where we set $M=1$.
 The contribution of the imaginary part is still small.
 The inset of each panel shows the existence of the maximum of $|V_{\rm D}|$.}
 \label{fig:Vmp_deep}
\end{figure}

First, we present the behavior of 
the real part of the potential ${\rm Re}(V_{\rm D})$,
the imaginary part ${\rm Im}(V_{\rm D})$ and 
the absolute value $|V_{\rm D}|$
for various non-dimensional spin parameters in Figs.~\ref{fig:Vmp070809}
and \ref{fig:Vmp_deep}.
The three panels in Fig.~\ref{fig:Vmp070809} 
are for  $q=0.7$ (left), $0.8$ (center) and $0.9$ (right).
We find that the contribution of the imaginary part is small even for $q=0.9$.
On the other hand, the three panels in Fig.~\ref{fig:Vmp_deep}
shows the potential for $q=0.93$ (left), $0.95$ (center) and $0.97$ (right).
Again, the contribution of the imaginary part is small,
but we see another peak around $r^*/M \sim 5$ in the $q=0.97$ case
(in practice, we also see another peak in the $q=0.95$ case
outside the figure).
This peak grows for larger $q$ and 
the height of the peak becomes dominant in the $q=0.98$ case.
Therefore, we restrict our analysis up to $q=0.97$ which is appropriate
since the recent numerical relativity results
suggest $q_f \lesssim 0.95$~\cite{Scheel:2014ina}.

In Table~\ref{tab:radii}, we summarize 
the event horizon radius $r_+$,
the peak location ($r_{\rm peak}$) of the absolute value of the potential $V_{\rm D}$,
and the location which satisfies $dV_{\rm D}/dr^*=0$
in the complex radius plane, $r_{dV_{\rm D}/dr^*=0}$.
The differences between the real part of $r_{dV_{\rm D}/dr^*=0}$
and $r_{\rm peak}$ are small,
and also the imaginary part of $dV_{\rm D}/dr^*=0$ is small.
In the same table, we also show the WKB result of ${\rm Im}(\omega)M$
and the solid angle of a sphere of $r_{\rm peak}$ soaking in the ergoregion ($4\pi C$)
estimated as
\bea
C= \frac{1}{2}\int_{\theta_{\rm m}}^{\pi-\theta_{\rm m}} \sin \theta d\theta \,,
\label{eq:C}
\eea
where $\theta_{\rm m}$ is calculated
by $r_{\rm peak}=M(1+\sqrt{1-q^2\cos^2\theta_{\rm m}})$.
The timelike Killing vector field of the Kerr space-time
becomes spacelike in the ergoregion,
and the region is coordinate invariant.
Here, we have introduced this $C$ as an estimator
which is less dependent on the coordinates,
while the radial coordinate is variant.

\begin{table}[!ht]
\caption{The event horizon radius, the peak location of the absolute value
of the potential $V_{\rm D}$,
and the location of $r$ which satisfies $dV_{\rm D}/dr^*=0$ in the complex radius plane.
We also show the WKB result of ${\rm Im}(\omega)M$.
The solid angle of a sphere of $r_{\rm peak}$ soaking in the ergoregion ($4\pi C$)
is estimated by Eq.~\eqref{eq:C}.
}
\label{tab:radii}
\begin{center}
\begin{tabular}{|c|c|c|c|c|c|}
\hline
$q$ & $r_+$/M & $r_{\rm peak}$/M & $r_{dV_{\rm D}/dr^*=0}/M$
& WKB Im($\omega$)M & $C$ \\
\hline
0.7 & 1.7141 & 1.9699 &
$1.9941 + 0.15560i$ & -0.082273 & 0.34786
\\
0.8 & 1.6 & 1.7585 &
$1.7852 + 0.15697i$ & -0.076730 & 0.81459
\\
0.9 & 1.4359 & 1.4969 &
$1.5169 + 0.13098i$ & -0.066175 & 0.96423
\\
0.91 & 1.4146 & 1.4664 &
$1.4850 + 0.12542i$ & -0.064379 & 0.97206
\\
0.92 & 1.3919 & 1.4348 &
$1.4517 + 0.11906i$ & -0.062330 & 0.97883
\\
0.93 & 1.3676 & 1.4018 &
$1.4171 + 0.11181i$ & -0.059964 & 0.98465
\\
0.94 & 1.3412 & 1.3670 &
$1.3808 + 0.10356i$ & -0.057193 & 0.98960
\\
0.95 & 1.3122 & 1.3301 &
$1.3425 +0.094182i$ & -0.053913 & 0.99363
\\
0.96 & 1.28 & 1.2901 &
$1.3018 +0.083455i$ & -0.049916 & 0.99687
\\
0.97 & 1.2431 & 1.2454 &
$1.2583 +0.070499i$ & -0.044759 & 0.99940
\\
\hline
\end{tabular}
\end{center}
\end{table}

\begin{figure}[!ht]
\begin{center}
 \includegraphics[width=0.48\textwidth,clip=true]{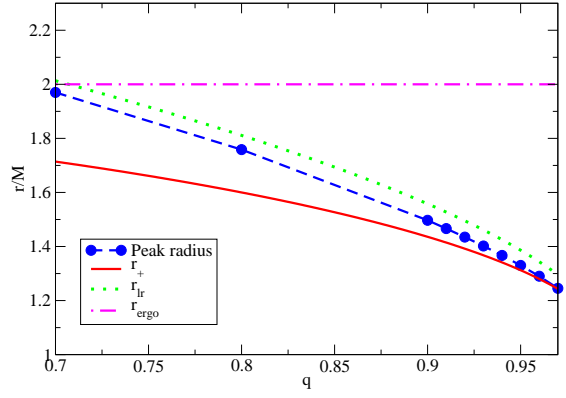}
\end{center}
 \caption{The location of the maximum of the absolute value of the potential
 $|V_{\rm D}|$ with ($\ell=2,\,m=2$).
 We also show the event horizon $r_{+}/M=1 + \sqrt{1-q^2}$,
 the inner light ring radius $r_{\rm lr}/M=2(1+\cos((2/3)\cos^{-1}(-q)))$,
 and the equatorial radius of the ergoregion $r_{\rm ergo}/M=2$
 evaluated for various spin parameters $q=a/M$.}
 \label{fig:location_mp}
\end{figure}

The peak location ($r_{\rm peak}$) of the absolute value of the potential $V_{\rm D}$
in Table~\ref{tab:radii} is plotted in Fig.~\ref{fig:location_mp}.
In this figure, we also show the event horizon radius $r_{+}$, 
the inner light ring radius $r_{\rm lr}$, 
and the equatorial radius of the ergoregion $r_{\rm ergo}$.

\begin{figure}[!ht]
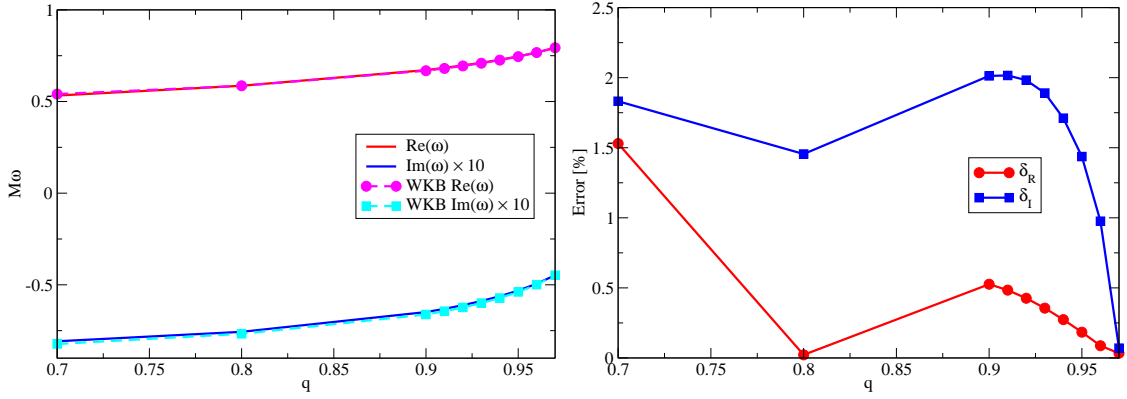

\begin{center}
 \includegraphics[width=0.48\textwidth,clip=true]{./frequencies_mp}
 \includegraphics[width=0.48\textwidth,clip=true]{./error_mp}
\end{center}
 \caption{(Left) The real and imaginary parts of the fundamental ($n=0$)
 QNM frequencies with $V_{\rm D}(\ell=2,\,m=2)$ evaluated for various spin parameters $q=a/M$.
 The exact frequencies ${\rm Re}(\omega)$ and ${\rm Im}(\omega)$
 are from Ref.~\cite{Berti:2005ys,BertiQNM}.
 (Right) Absolute value of relative errors for the real and imaginary part
 of the QNM frequencies with $V_{\rm D}(\ell=2,\,m=2)$,
 $\delta_{\rm R}=|({\rm WKB}~{\rm Re}(\omega))/{\rm Re}(\omega)-1|$
 and $\delta_{\rm I}=|({\rm WKB}~{\rm Im}(\omega))/{\rm Im}(\omega)-1|$
 between the exact value and that of the WKB approximation.}
 \label{fig:frequencies_mp}
\end{figure}

The real and imaginary parts of the $n=0$ QNM frequencies
are shown in the left panel of Fig.~\ref{fig:frequencies_mp}, respectively.
Here, we present the WKB result and the exact frequencies
given in Ref.~\cite{Berti:2005ys}.
We find that the errors are quite small in this spin range
(see the right panel of Fig.~\ref{fig:frequencies_mp}).

\section{Discussions}

In our previous paper~\cite{Nakamura:2016gri},
we assumed that the remnant BH spin
$q_f=0.7$ to compute the detection rate.
When we treat BHs with $q_f=0.95$, 
two effects from the QNM frequency arise,
that is, the (real part of) frequency of QNM increases
while the damping rate decreases. Therefore, the recalculation is needed
(note that the QNM amplitude (excitation) is also an important input).
However, the former effect will decrease the event rate
while the latter effect increases the event rate.
Therefore, the event rate will be more or less similar.
This means that the merging rate of Pop III
$\sim 30 M_\odot$--$30M_\odot$ BH binaries
with the signal-to-noise ratio $> 35$ needed
for the determination of QNMs~\cite{Nakano:2015uja} is 
roughly $0.17$--$7.2$~${\rm events~yr^{-1}~(SFR_p/(10^{-2.5}~M_\odot~yr^{-1}~Mpc^{-3}))}
\cdot (\rm [f_b/(1+f_b)]/0.33)$
where ${\rm SFR_p}$ and ${\rm f_b}$ are the peak value
of the Pop III star formation rate and the fraction of binaries,
respectively~\cite{Kinugawa:2014zha,Kinugawa:2015}.

\begin{figure}[!ht]
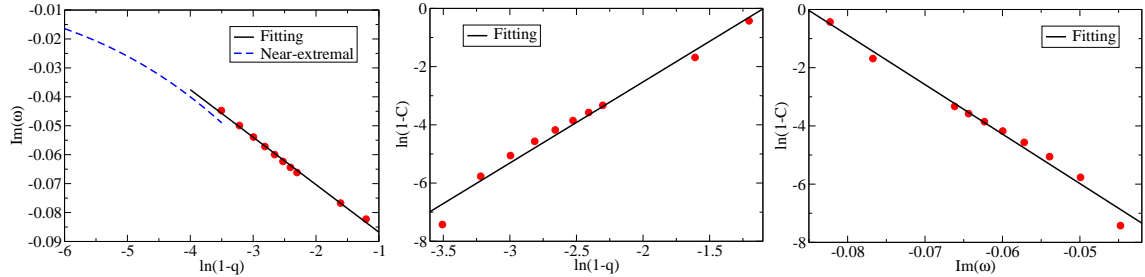

\begin{center}
 \includegraphics[width=0.325\textwidth,clip=true]{./fitting1}
 \includegraphics[width=0.32\textwidth,clip=true]{./fitting2}
 \includegraphics[width=0.32\textwidth,clip=true]{./fitting3}
\end{center}
 \caption{The red points are from Table~\ref{tab:radii},
 and the black lines show Eqs.~\eqref{eq:relations} obtained by linear-fitting.
 In the left panel, we also present Eq.~\eqref{eq:NEwI} by the blue dashed curve.}
 \label{fig:fitting123}
\end{figure}

As seen in Fig.~\ref{fig:location_mp}, 
the location of the maximum of the absolute value of the potential
is within the ergoregion for $q \gtrsim 0.7$.
It is noted that in Table~\ref{tab:radii}
the spin, the imaginary part of the QNM frequency,
and the solid angle of a sphere of $r_{\rm peak}$ soaking
in the ergoregion have simple relations, 
\bea
 M{\rm Im}(\omega) &=& -0.016416 \,\ln(1-q) - 0.1032 \,,
 \cr 
 \ln(1-C) &=& 2.7867 \,\ln(1-q) + 3.0479 \,,
 \cr
 \ln(1-C) &=& -169.92 \,M {\rm Im}(\omega) - 14.476 \,,
 \label{eq:relations}
\eea
in the range between $q=0.7$ and $0.97$ (see Fig.~\ref{fig:fitting123}).
Related to the existence of the event horizon, it is important to know
how far we can confirm the ergoregion by gravitational wave detection of
the QNMs. The above relations give us the direct relation between $C$
and the imaginary part of the observed QNM frequency.

Here, according to Hod~\cite{Hod:2008zz},
the imaginary part of the QNM frequency for near extremal Kerr BHs is
\bea
 M{\rm Im}(\omega) = - \frac{\sqrt{1-q^2}}{4(1+\sqrt{1-q^2})} \,,
 \label{eq:NEwI}
\eea
which is shown as the blue dashed curve
in the left panel of Fig.~\ref{fig:fitting123}.
The difference between Eqs.~\eqref{eq:relations} and \eqref{eq:NEwI}
can be considered as the next order effect of the near-extremal approximation.
We will clarify the physical meaning of Eqs.~\eqref{eq:relations} in a future work.

Finally, it is necessary to investigate QNMs in the case of $q >0.97$
that requires some high accuracy study (see e.g., Ref.~\cite{Cook:2014cta}).
Also, there are many excited states~\cite{Sasaki:1989ca}
which contribute to the QNM gravitational waves.
The detailed study is needed for a fully-integrated understanding of QNMs.

In conclusion, we have covered the QNM analysis of the Kerr BH
up to $q \sim 0.95$ in this paper.
This spin parameter $q \sim 0.95$ is expected by state-of-the-art
numerical relativity simulations of mergers of a equal-mass, highly spinning BH binary 
with aligned spins of the same magnitude $q=0.994$.
The detection of the QNM with $a/M \sim 0.95$
by the second generation gravitational wave detectors
would prove a sphere $99\%$ soaking in the ergoregion of the Kerr BH,
and such a  space-time region will give us
the opportunity to test Einstein's general relativity in strong gravity.

\section*{Acknowledgment}

~~This work was supported by MEXT Grant-in-Aid for Scientific Research
on Innovative Areas,
``New Developments in Astrophysics Through Multi-Messenger Observations
of Gravitational Wave Sources'', No.~24103006 (HN, TN, TT) and
by the Grant-in-Aid from the Ministry of Education, Culture, Sports,
Science and Technology (MEXT) of Japan, No.~15H02087 (TN, TT).


\end{document}